# Ambient PMU Data Based System Oscillation Analysis Using Multivariate Empirical Mode Decomposition

Shutang You

**Abstract**: Wide-area synchrophasor ambient measurements provide a valuable data source for real-time oscillation mode monitoring and analysis. This paper introduces a novel method for identifying inter-area oscillation modes using wide-area ambient measurements. Based on multivariate empirical mode decomposition (MEMD), which can analyze multi-channel non-stationary and nonlinear signals, the proposed method is capable of detecting the common oscillation mode that exists in multiple synchrophasor measurements at low amplitudes. Test results based on two real-world datasets validate the effectiveness of the proposed method.

## Keywords

Wide-area synchrophasor measurement; ambient oscillation mode identification; multivariate empirical mode decomposition (MEMD).

## 1. Introduction

Power system oscillation is a common phenomenon in interconnected power grids. Insufficient damping of inter-area oscillation may increase system risks and even cause failures [1]. Therefore, fast detection and analysis on inter-area oscillations are critical to activate proper oscillation damping controls to increase system reliability. Moreover, as the conditions of modern power girds vary more constantly and significantly with the increase of renewables, energy storage and other distributed resources, updating the oscillation information is becoming more important but also challenging.

Approaches to analyze the oscillation modes of power grids fall into two categories: model-based methods and measurement-based methods. Model-based methods analyze system oscillation based on detailed system dynamic models using eigenvalue analysis approaches. It is becoming difficult for system operators to use model-based approaches due to many factors, such as the computation burden of large-scale dynamic simulations, changing environments, information privacy issues and parameter inaccuracy (e.g. inadequate modeling of loads) [2]. Applying synchrophasor measurement technology, Wide-Area Measurement Systems (WAMSs) provide a powerful tool to monitor and analyze the dynamics of interconnected power grids [3-5]. Since large disturbances are rare and usually destructive in modern power grids, system event data are far from enough for real-time estimation of system oscillation modes. It is necessary to provide oscillation information to system operators under normal operation conditions. Studies on ambient synchrophasor measurements show that there is a constant level of noise caused by load variations or other environmental disturbances ae transmission level [6] and the distribution level [3]. These ambient sychrophasor measurements have been recently used as a data source to extract real-time inter-area

Email: syou3@utk.edu.

oscillation information.

Since ambient measurements were first used in Ref. [7] to analyze the electromechanical oscillation, multiple approaches have been developed based on various signal processing techniques. There are two main categories of methods for oscillation analysis based on ambient and event data: transfer function based methods and subspace methods. Transfer-function-based methods directly estimate mode shape through treating measurements as system outputs. Typical transfer function based methods include Fourier transform [8-10], the Prony's method [11], the Matrix-Pencil method [12], Empirical Mode Decomposition (EMD) [13], the Yule-Walker method [14], and the singular value decomposition method [15], etc. To increase analysis efficiency, Ref. [9] adopted a FFT-based distributed optimization method to select the dominant measurement channels for estimating each oscillation mode based on ambient data. Ref. [16] proposed a two-step method which comprised of independent component analysis and random decrement to estimate the oscillation mode. Different from transfer-function-based methods, subspace methods obtain the oscillation mode information through identifying the system state space model using the measurements [7]. Typical subspace methods include the Canonical Variate Algorithm [17], the N4SID algorithm [18], and the autoregressive moving average block-processing method [19]. Recently, Ref. [20] proposed the robust recursive least square algorithm to analyze measurement data. Ref. [21] improved this method by proposing a regularized robust recursive least square method.

Existing transfer function based methods have limitations in one or two of the following aspects. *a) The capability to analyze drifting, non-stationary signals and provide localized results.* Oscillation may drift frequently in ambient measurements [22]. For example, a high damping local oscillation mode may stimulate an inter-area oscillation mode with a pseudo negative damping ratio within a short duration. Existing methods may not be able to find the appropriate time window for analyzing this inter-area oscillation due to this drifting [23]. In addition, some measurements are mixed with fluctuations unrelated to oscillations, such as the frequency fluctuations caused by normal system operation and regulations [24]. These non-stationary measurements usually lead to difficulties to accurately analyze the subtle oscillation modes [25]. For example, the Prony's method requires the signal to be zero-mean and stationary, which may result to difficulties to analyze signals with high amplitude trends [19]. *b) The capability to analyze multi-channel signals.* Most existing methods can only process one signal so they have to analyze multiple signals in a separate way. they usually require measurements from critical devices, which has good observability (e.g., branch flow and bus frequency) on certain inter-area oscillation modes. If the measurement that has good oscillation observability is not available in some areas, single-channel methods may not be able to provide oscillation information in these areas. Therefore, there is a need of developing multi-channel methodology to extract oscillation information utilizing high-noise measurements, which individually have less observability on oscillation.

The aim of this study is to introduce multi-channel EMD based methods: Bivariate EMD (BEMD), Trivariate EMD (TEMD), and Multivariate EMD (MEMD), to facilitate identifying inter-area oscillation mode using multi-channel ambient synchrophasor measurements. Particularly, MEMD is investigated in details using real-world ambient measurements. The test results show that MEMD has better performance than typical oscillation identification methods. For convenience, wide-area frequency measurements are adopted for illustration and analysis in the following sections. The introduced methods and descriptions are also applicable to other wide-area measurement data such as instantaneous real power, voltage angle, and current.

The rest of this paper is organized as follows. Section 2 describes the original form of the EMD method. Section 3 expands it to bivariate, trivariate, and multivariate empirical mode decomposition to identify oscillation. Section 4 presents two test cases to verify the proposed methods. Conclusions are provided in Section 5.

## 2. Background — Empirical Mode Decomposition Based Oscillation Identification

EMD was developed for analyzing non-stationary signals that commonly exist in many science and engineering fields [1]. Due to its data-driven nature and the strong capability in analyzing non-stationary signals to provide information on localized amplitudes and frequencies, EMD has been proved to be effective for time-frequency analysis in various areas, such as power quality assessment [26], biomedical signals [27], mechanical signals [28], and geographical signals [29]. For oscillation identification, EMD has been applied to analyze transient measurements [13].

In EMD-based oscillation identification, a frequency measurement signal can be viewed as a linear combination of short-term frequency fluctuation components and long-term frequency trends [30]. Short-term frequency fluctuations are defined as the evolution feature of frequency measurements $f(t)$ between local frequency maxima and minima. Subtracting this fast fluctuation component, which is denoted by $\tilde{f}_1(t)$, from $f(t)$, one can identify the "slower" frequency trend $\bar{\bar{f}}_1(t)$ that supports the short-term frequency fluctuation component, so that

$$f(t) = \bar{\bar{f}}_1(t) + \tilde{f}_1(t) \tag{1}$$

where $\tilde{f}_1(t)$ is termed an intrinsic mode function (IMF). As raw measurements have noises or measurement errors, $\tilde{f}_1(t)$ represents these high-frequency elements. $\bar{\bar{f}}_1(t)$ is still oscillatory after subtracting the fast fluctuation component $\tilde{f}_1(t)$ from $f(t)$. The same decomposition method can be applied to $\bar{\bar{f}}_1(t)$ as $\bar{\bar{f}}_1(t) = \bar{\bar{f}}_2(t) + \tilde{f}_2(t)$. After recursive decompositions, the representation of the original frequency measurement $f(t)$ becomes

$$f(t) = \bar{\bar{f}}_M(t) + \sum_{m=1}^{M} \tilde{f}_m(t) \tag{2}$$

where $\tilde{f}_m(t)$ is the *m*th IMF of frequency measurements and $\bar{\bar{f}}_M(t)$ is the residual frequency. Each iterative step for decomposing the frequency signal into one IMF has the iterative procedures as shown in Table 1 [2].

Table 1. Iteration steps of univariate EMD to generate each IMF

| Step | Procedure |
|---|---|
| 1) | Find all frequency extrema of $f(t)$, which are denoted by $\bar{f_i}$ (maxima) and $\underline{f_i}$ (minima). |
| 2) | Construct two time-series frequency signals from maxima $\bar{f_i}$ and minima $\underline{f_i}$, respectively, through cubic-spline interpolation. The two constructed frequency signals $\bar{\bar{f}}_{max}$ and $\bar{\bar{f}}_{min}$ form an envelope. |
| 3) | Computing the mean of the envelop: $\bar{\bar{f}}(t) = (\bar{\bar{f}}_{max} + \bar{\bar{f}}_{min})/2$. |
| 4) | Subtracting the envelop mean to obtain a fast fluctuation components $\tilde{f}(t) = f(t) - \bar{\bar{f}}(t)$. |
| 5) | Continue to sift $\tilde{f}(t)$ using Step 1) to 4) until it meets the IMF criterion. Afterwards, get the residue $f(t) - \tilde{f}(t)$ and return to Step 1) to decompose the next IMF. |

In Step 5), the IMF criterion to stop the sifting process for generating one IMF is that the normalized square deviation of two consecutive sifted signals is smaller than a threshold.

$$\sum_{t=1}^{T} \left[ \frac{\left| \tilde{f}_{m,l+1} - \tilde{f}_{m,l} \right|^2}{\tilde{f}_{m,l}{}^2} \right] \leq \bar{D}_{\text{EMD}} \tag{3}$$

where $l$ and $l$+1 denote two successive sifting operations. The typical value of $\bar{D}$ is between 0.2 and 0.3 [30]. The stopping criterion for the outer iterative loop (decomposing IMFs) is that the residue $\bar{\bar{f}}_K(t)$ becomes a monotonic signal, from which no more IMFs can be extracted.

Since the original EMD can only process the real-value (univariate) time series signal, existing work on oscillation identification are limited to applying classical EMD or Ensemble EMD (EEMD, an modified version of EMD) on simulation data or transient measurements [13, 31]. As described in Section 1, ambient measurements may contains valuable information such as common oscillation components that reflect inter-area oscillation modes. This information may be difficult to extract using univariate EMD due to low signal/noise ratio environments.

# 3. Methodology — Multivariate Empirical Mode Decomposition Based Ambient Oscillation Mode Identification

## 3.1 Bivariate/Trivariate Empirical Mode Decomposition for Oscillation Identification

The bivariate (complex) EMD (BEMD) [32] and the trivariate EMD (TEMD) [33] enhanced the capability of identifying synchronous behaviors of bivariate and trivariate signals. In many fields, BEMD and TEMD has been proved to be able to determine common frequency components through simultaneous decomposition of two or three signals, such as equipment condition monitoring [34] and biomedical signal analysis [35]. Since inter-area oscillations typically happen in two or three areas, if the measurements from two or three areas are available, it is possible to apply BEMD and TEMD to multiple channels of frequency ambient measurements to improve oscillation identification. Based on their basic algorithm formulation [32, 33], the iterative procedures to extract each oscillation component based on BEMD and TEMD are developed in Table 2 and Table 3, respectively.

Table 2. Iterative procedures of BEMD for generating each IMF

| Step | Procedure |
|---|---|
| 1) | Use two frequency measurements $f_A$ and $f_B$ in area A and B to construct the complex frequency measurement vector $\boldsymbol{f}_{\text{A}-\text{B}} = f_{\text{A}} + i * f_{\text{B}}$. |
| 2) | Construct $N$ directions for projection of the frequency vector. These directions are denoted by $\varphi_k = 2k\pi/N\,, 1 \leq k \leq N$. |
| 3) | Project the complex frequency measurement vector $\boldsymbol{f}_{\text{A}-\text{B}}$ on direction $\varphi_k$ : $P_{\varphi_k}(t) = \text{Re}\left(e^{-i\varphi_k}x(t)\right)$. |
| 4) | Extract the time of the maxima in $P_{\varphi_k}(t)$. The time instants at maxima are denoted by $\{t_j^k\}$. |
| 5) | Conduct cubic-spline interpolation based on the point set $\left[t_j^k, \boldsymbol{f}_{\text{A}-\text{B}}\left(t_j^k\right)\right]$ to get the complex frequency envelope curve in the direction $\varphi_k$, which is denoted by $\boldsymbol{e}_{\varphi_k}$. Update the direction index from $k$ to $k$+1. If $k$<$K$, return to Step 3). Otherwise, continue the next step. |
| 6) | Calculate the complex frequency envelop mean: $\bar{\bar{\boldsymbol{f}}}(t) = 2/N \cdot \sum_k \boldsymbol{e}_{\varphi_k}(t)$. |
| 7) | Subtract $\bar{\bar{\boldsymbol{f}}}(t)$ from the complex frequency measurement vector $\boldsymbol{f}_{A-B}$, so that $\tilde{\boldsymbol{f}}(t) = \boldsymbol{f}_{\text{A}-\text{B}}(t) - \bar{\bar{\boldsymbol{f}}}(t)$. |
| 8) | Continue to sift $\tilde{\boldsymbol{f}}(t)$ using Step 1) to 7) until it meets the IMF criterion. Afterwards, get the residue $\boldsymbol{f}_{\text{A}-\text{B}}(t) - \tilde{\boldsymbol{f}}(t)$ and return to step 1 to decompose the next IMF. |

Table 3. Iterative procedures of TEMD for generating each IMF

| Step | Procedure |
|---|---|
| 1) | Using three frequency measurements $f_{\text{A}}$, $f_{\text{B}}$, and $f_{\text{C}}$ from three areas, construct the trivairate quaternion frequency signal denoted by $\boldsymbol{f}(t)$. |
| 2) | Calculate the projection of $\boldsymbol{f}(t)$, i.e., $p_{\theta_k}^{\varphi_n}$, where $\theta_k = k\pi/K$ and $\varphi_n = n\pi/N$, where $k = 1, \ldots, K \; and \; n = 1, \ldots, N$. |
| 3) | Extract the locations $\{(t_k^n)_i\}$ of the maxima of $p_{\theta_k}^{\varphi_n}(t)$ for eac1h $k$ and $n$. |
| 4) | Conduct cubic-spline interpolation on the extrema point set $[(t_k^n)_i, \boldsymbol{f}(t_k^n)_i]$ to get the frequency envelope curve in the direction $\{\theta_k, \varphi_n\}$, which is denoted by $\boldsymbol{e}_{\theta_k}^{\varphi_n}$ for each $k$ and $n$. |
| 5) | Calculate the complex frequency envelop mean: $\bar{\bar{\boldsymbol{f}}}(t) = \frac{1}{KN} \cdot \sum_k \sum_n \boldsymbol{e}_{\theta_k}^{\varphi_n}$. |
| 6) | Subtract $\bar{\bar{\boldsymbol{f}}}(t)$ so that $\tilde{\boldsymbol{f}}(t) = \boldsymbol{f}(t) - \bar{\bar{\boldsymbol{f}}}(t)$. |
| 7) | Continue to sift $\tilde{\boldsymbol{f}}(t)$ using Step 1) to 6) until it meets the IMF criterion. Afterwards, return to Step 1) to decompose the next IMF. |

The IMF criterion to stop the IMF sifting are extensions of (3) to multiple signals, For example, the IMF criterion for the TEMD case is shown in (4).

$$\frac{1}{3}\sum_{A,B,C}\sum_{t=1}^{T}\left[\frac{\left|\tilde{f}_{m,l+1}-\tilde{f}_{m,l}\right|^2}{\tilde{f}_{m,l}{}^2}\right] \leq \bar{D}_{\text{TEMD}} \tag{4}$$

The criterion to stop decomposing the next IMF is that all residual signal projections have less than two extrema (for BEMD) or three extrema (for TEMD).

### 3.2 Multivariate Empirical Mode Decomposition for Oscillation Identification

Multivariate EMD (MEMD) was developed for analyzing complex, nonlinear, and dynamic signals [36]. The simultaneous analysis of multi-channel signals using MEMD has been proved to be capable to analyze multichannel signals in a synchronized approach [37]. This synchronized approach helps to identify the oscillation modes that exists in multiple channels with low amplitudes, which may be neglected in mono-channel analysis [38]. This feature makes MEMD a competitive candidate method for oscillation mode identification based on wide-area ambient measurements.

The essence of EMD-based methods is based on local maxima and minima information, but local extrema definition are not obvious for multiple-channel frequency signals [36]. Therefore, the main difficulty of extending EMD to MEMD for frequency analysis is to generate the frequency envelops based on local extrema. Assuming $\{\boldsymbol{f}_N(t)\}_{t=1}^{T} = \{f_1(t), f_2(t), \dots, f_N(t)\}$, the main steps of applying MEMD [36] for oscillation mode identification are described in Table 4.

Table 4. Iterative procedures of MEMD for generating each IMF

| Step | Procedure |
|---|---|
| 1) | Generate a set of angles $\theta^k\}_{k=1}^{K}$ on a sphere that has ($n$−1) dimensions to denote $K$ projection directions. Create the direction vectors denoted by $X^{\theta_k}\}_{k=1}^{K}$ based on these angles. |
| 2) | Project the frequency measurements $\boldsymbol{f}_N(t)$ for all $K$ directions. The projections are denoted by $p^{\theta_k}(t)\}_{k=1}^{K}$. |
| 3) | Find the maxima of the projections. Give $\left\{t_i^{\theta_k}\right\}$ as the time instants at the maxima of projections $p^{\theta_k}(t)\}_{k=1}^{K}$. |
| 4) | Use maxima $\left[t_i^{\theta_k}, \boldsymbol{f}\left(t_i^{\theta_k}\right)\right]$ and cubic-spline interpolation to obtain the envelop curve in each direction. The envelop curves for all $K$ directions are denoted by $\boldsymbol{e}^{\theta_k}\}_{k=1}^{K}$. |
| 5) | Calculate the envelop mean $\bar{\bar{\boldsymbol{f}}}(t) = \frac{1}{K}\sum_{k=1}^{K} \boldsymbol{e}^{\theta_k}(t)$. Subtract $\bar{\bar{\boldsymbol{f}}}(t)$ so that $\tilde{\boldsymbol{f}}(t) = \boldsymbol{f}(t) - \bar{\bar{\boldsymbol{f}}}(t)$. |
| 6) | Continue to sift $\tilde{\boldsymbol{f}}(t)$ using Step 1) to 5) until it meets the IMF criterion. Afterwards, return to Step 1) to decompose the next IMF. |

The IMF criterion in Step 6) to stop the sifting iteration is shown in (5).

$$\frac{1}{N}\sum_{n=1}^{N}\sum_{t=1}^{T}\left[\frac{\left|\tilde{f}_{m,l+1}-\tilde{f}_{m,l}\right|^2}{\tilde{f}_{m,l}^{\ 2}}\right] \leq \bar{D}_{\text{MEMD}} \tag{5}$$

The outer iteration loop for generating the next IMF terminates when signal projections $p^{\theta_k}(t)\}_{k=1}^{K}$ have less than three extrema.

Using MEMD, multi-channel ambient frequency measurements are analyzed as an $n$ dimensional matrix and decomposed into several IMFs components based on the iterative process in Table 4. Each IMF contains the localized frequency, amplitude, and phase angle information for each frequency signal and oscillation mode. Dominant oscillation modes can be identified based on the energy of each IMF given by

$$E_{k,\infty} = \sum_{t=1}^{T} \sum_{n=1}^{N} \left[\tilde{f}_k(t)[n]\right]^2 \quad (6)$$

In some situations, the measurements may have high-magnitude trends that are extracted as high-energy IMFs. These high magnitude trends can be easily excluded from inter-area oscillations as the IMF for all signals have very close phase angles. Additionally, it is also easy to distinguish these IMFs from their frequencies, which are usually lower than typical interarea oscillation frequencies.

Through the Hilbert Transform [30], the localized frequency and amplitude values for each IMF and each signal can be obtained as

$$f_H(t) = \frac{P}{\pi} \int_{-\infty}^{+\infty} \frac{f(\tau)}{t-\tau} d\tau \quad (7)$$

$$z(t) = f(t) + i f_H(t) = a(t) e^{i\varphi(t)} \quad (8)$$

where $a(t)$ is the localized amplitude obtained by $a(t) = \left[f^2(t) + {f_H}^2(t)\right]^{1/2}$. $\varphi(t)$ is the localized angle calculated by $\varphi(t) = \arcsin(f_H(t)/a(t))$. The instantaneous frequency for all IMFs and signals can be obtained as $\omega(t) = d\varphi(t)/dt$. The joint instantaneous frequency and amplitude of one IMF denote the frequency and amplitude for a particular oscillation mode considering all signals at the system level, and they are obtained by

$$f_{\mathrm{IMF},l}(t) = \frac{\sum_{n=1}^{N}\left(a_{\mathrm{IMF},l,n}(t) \cdot f_{\mathrm{IMF},l,n}(t)\right)}{\sum_{n=1}^{N} a_{\mathrm{IMF},l,n}(t)} \quad (9)$$

$$a_{\mathrm{IMF},l}(t) = 1/N \cdot \sum_{n=1}^{N} a_{\mathrm{IMF},l,n}(t) \quad (10)$$

The joint instantaneous frequency and amplitude for each IMF can be graphed for visualization in the time domain. This information can help operators understand real-time oscillation information, as well as resonance-stimulated oscillations based on the oscillation occurrence sequence. In addition, the computational complexity of EMD-based algorithms are proved to be equivalent to Fast Fourier Transform (FFT) [39], indicating MEMD is suitable for online oscillation identification.

## 4. Case Studies Based on FNET/GridEye Measurements

This section shows two cases to investigate the features of TEMD and MEMD for oscillation identification based on FNET/GridEye ambient measurements [40-64]. As a WAMS at the distribution level, the FNET/GridEye system is capable of monitoring the power grid with high dynamic accuracy [65, 66]. A frequency disturbance recorder (FDR) can measure power grid voltage, angle, and frequency at the 120V outlets. These highly accurate synchrophasor measurements at transmitted across the Internet at a 10Hz report rate (i.e. 10 frequency samples per second) and collected by the main server located at the University of Tennessee. Table I shows the description on the two cases using FNET/GridEye ambient measurements.

Table 5. Information on the study cases

| Case # | Measurement information | Test purpose |
|---|---|---|
| 1 | FNET/GirdEye ambient frequency measurements in Europe | a) Test the capability of TEMD in identifying oscillation modes based on ambient measurements<br>b) Compare TEMD with classical EMD |
| 2 | FNET/GirdEye ambient and event measurements in U.S. Eastern Interconnection | a) Test the capability of MEMD in identifying oscillation modes using high-noisy ambient data<br>b) Test the capability of MEMD in analyzing non-stationary frequency measurements during events<br>c) Verify the consistency of identified oscillation modes using ambient and event measurements<br>d) Compare MEMD with typical methods |

### 4.1 European Ambient Frequency Measurements

The first case uses three channels of ambient frequency measurements at three locations in the European power grid as shown in Fig. 1. Applying TEMD to the measurements, Fig. 2 to Fig. 4 show the IMFs in the TEMD results. IMF 1 and IMF 2 show the high frequency components that are caused by measurement errors and noises. IMF 3 represents a 1.0 Hz local oscillation mode in the Turkey power grid. IMF 4 and IMF 7 have small amplitudes and don’t show dominant oscillations’ information. IMF 8 and IMF 9 contain high magnitudes of variations, reflecting long-term frequency fluctuations resulting from governor responses and automatic generation control. Based on the energy function (6) and IMF amplitudes, IMF 5 and IMF 6 can be as identified dominant inter-area oscillation modes, whose frequencies are 0.30 Hz and 0.15 Hz, respectively. It can be seen that the observed damping ratio of the two oscillation modes vary with time (some even have negative damping) because of constant small disturbances occurred in the ambient environments. Fig. 5 shows the joint instantaneous frequency based on Hilbert spectral analysis.

As a comparison, Fig. 6 shows IMF 4 to IMIF 6 using the classic EMD method. It can be seen that the two inter-area oscillation modes are mixed up with each other in IMF 4, thus unable to be identified using EMD. From this comparison, it can be noticed that if the measurements in one area has high noises, MEMD could extract possible oscillation information from the noisy signals in this area aided by oscillation information from other areas.

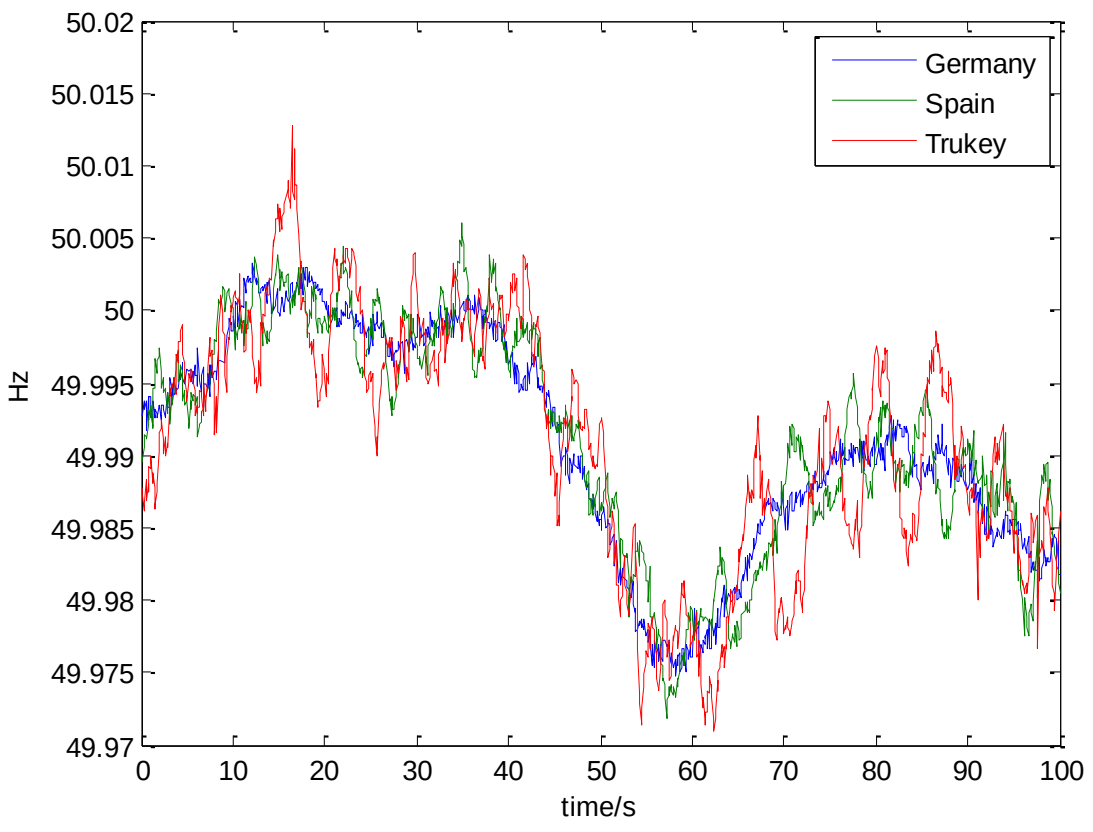

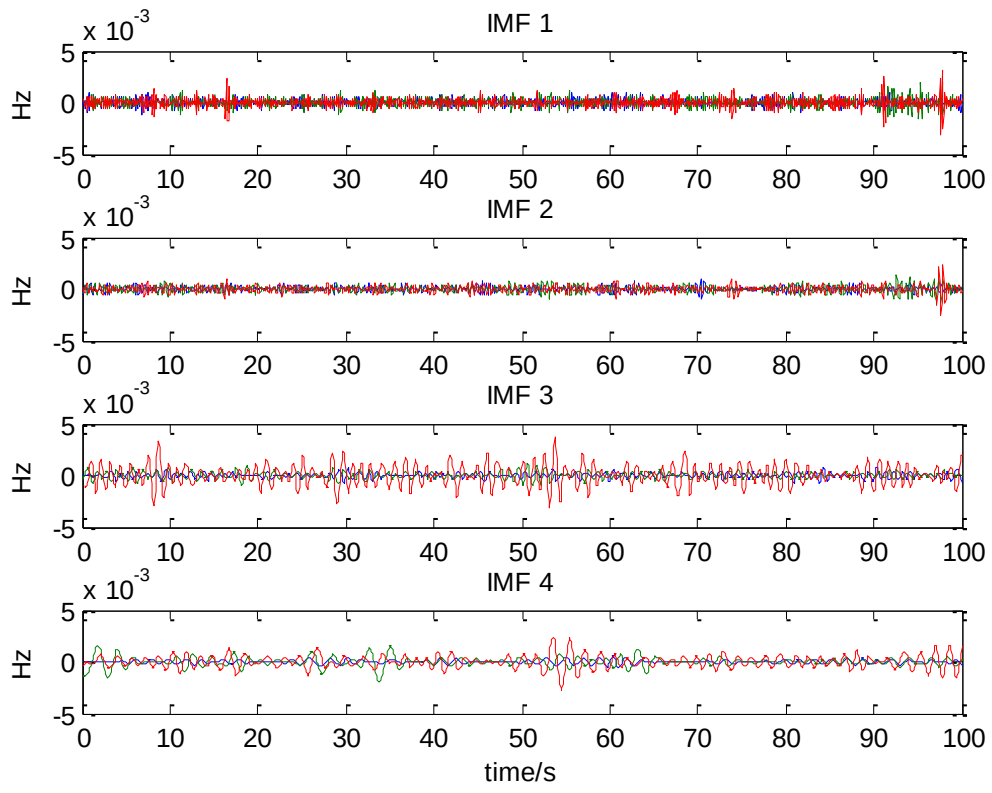

Fig. 1. The European case - ambient frequency measurements with obvious oscillation

Fig. 2. IMF 1-IMF 3 of the European case

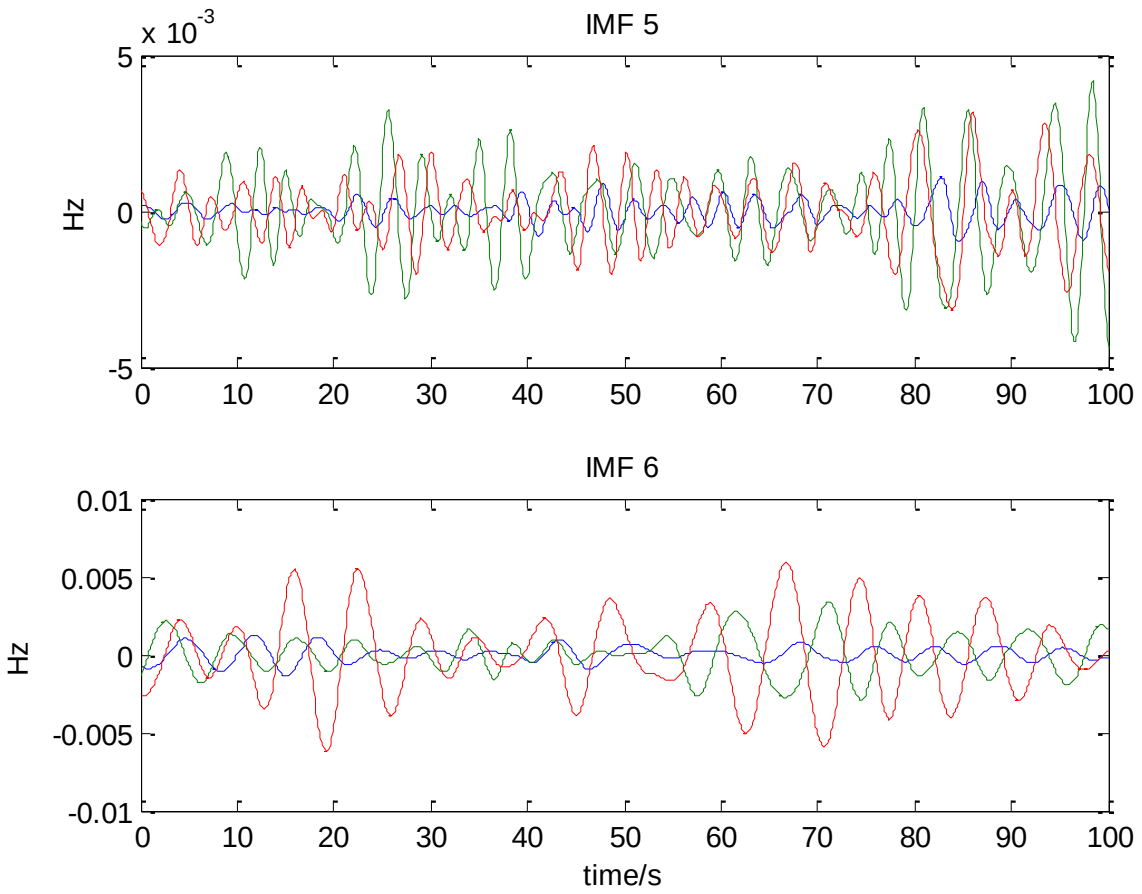

Fig. 3. IMF 4-IMF 6 of the European case

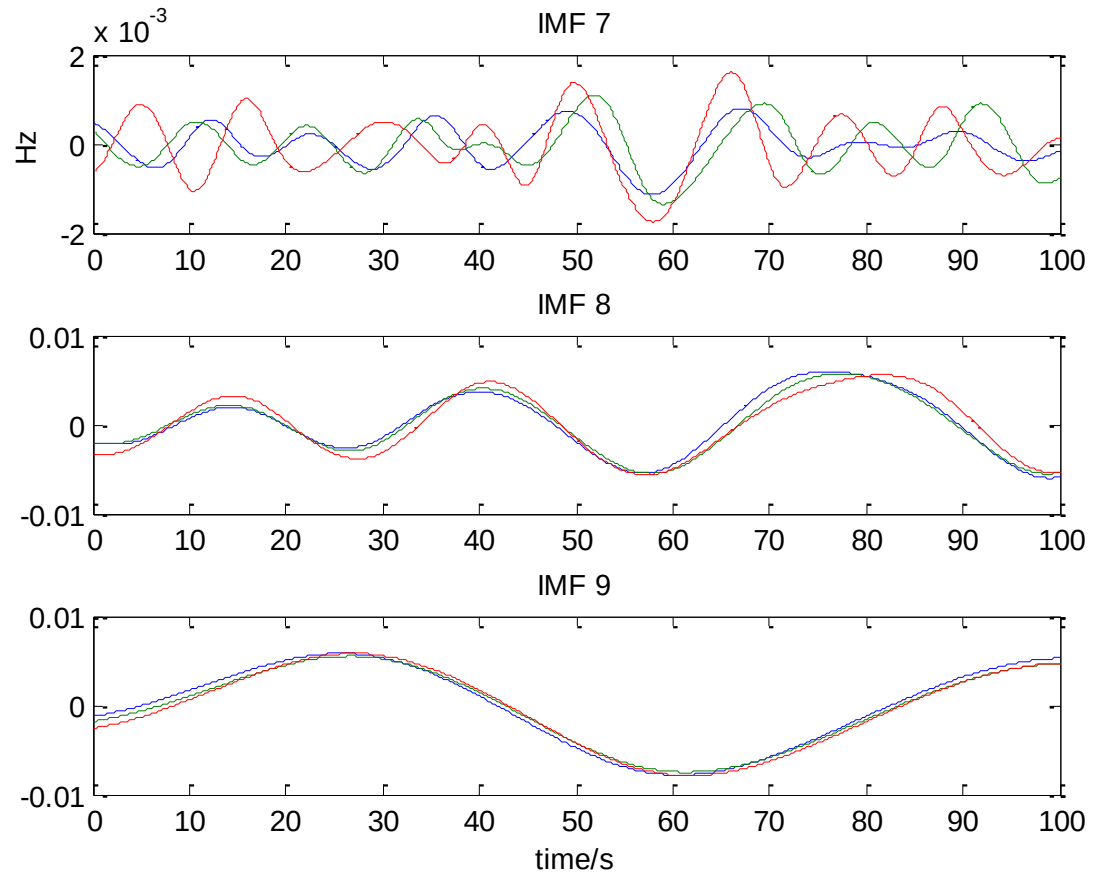

Fig. 4. IMF 7-IMF 9 of the European case

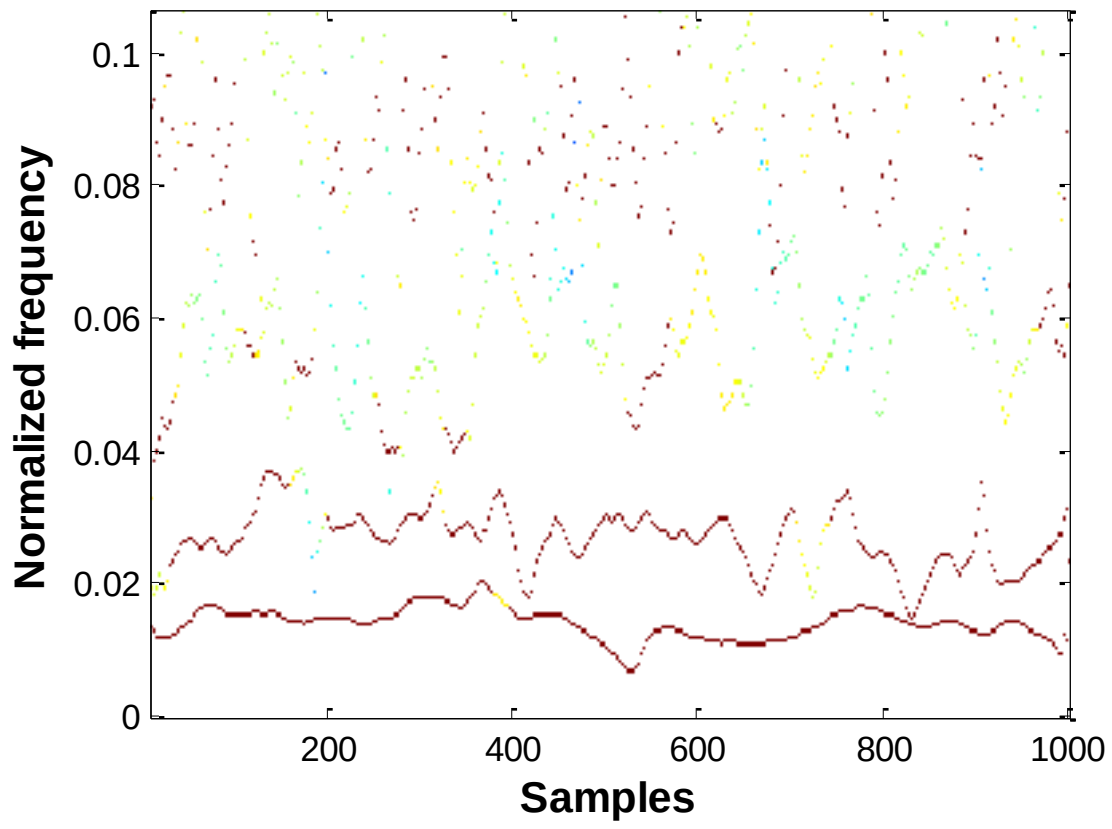

Fig. 5. The Hilbert spectrum (instantons frequency) of IMF 3 to IMF 6 in the European grid ambient measurements

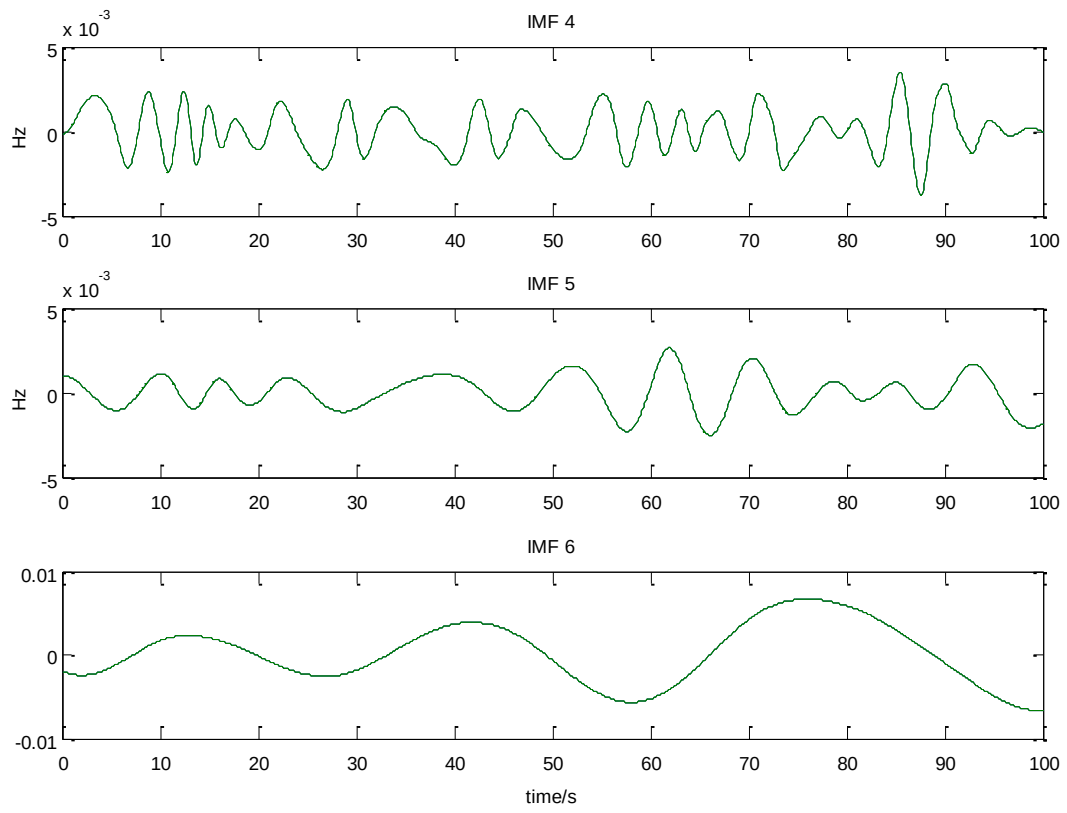

Fig. 6. IMF 4 to IMF 6 obtained by EMD

## 4.2 EI Ambient and Event Frequency Measurements

The second case analyzes ambient measurements and the measurements of a generation trip event in the U.S. Eastern Interconnection (EI) system. The deployment of measurement devices of FNET/GridEye in the EI system is shown in Fig. 7. This case aims to verify that the oscillation mode identified using the ambient data can be actually observed in event transients. Frequency measurements from 12 locations around the event occurrence time are shown in Fig. 8 . The studied time window is 5 minutes, during which the generation trip event happened at 125s.

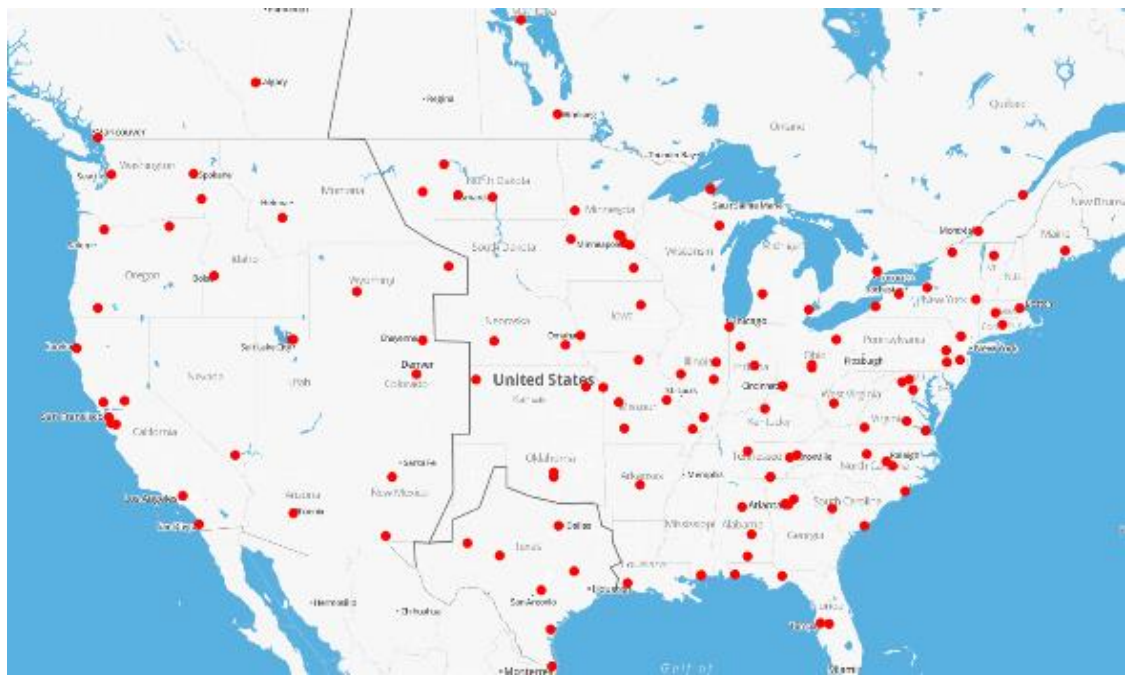

Fig. 7. FDR deployment in the North America

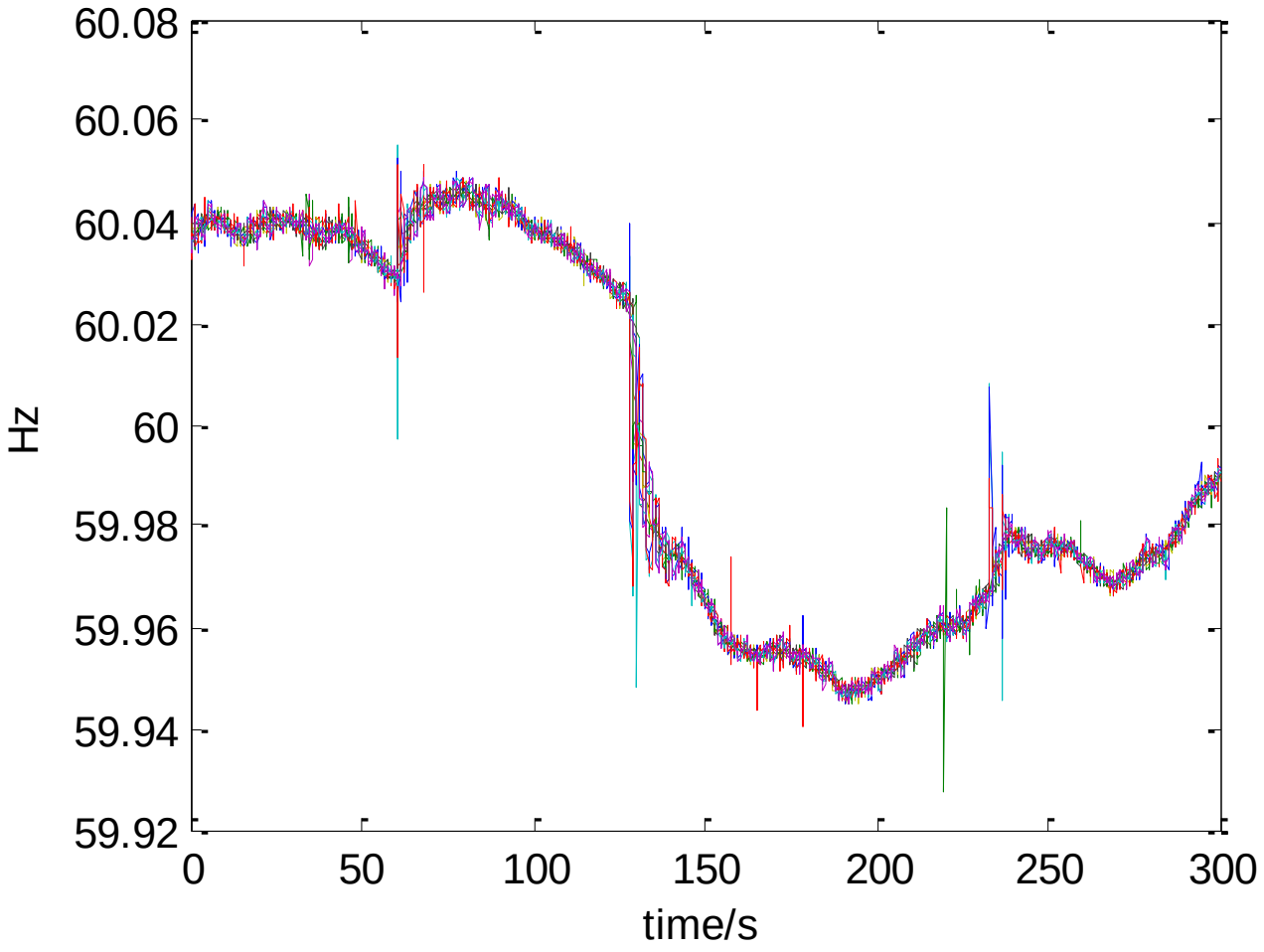

Fig. 8. Frequency measurements of a generation trip event in EI

Fig. 9 to Fig. 11 show the MEMD decomposition result using the frequency measurements. Among the obtained IMFs, IMF 1 to IMF 4 shows the high-frequency components caused by local system noises, measurement errors, and local oscillations with high damping ratios. Observed from its frequency and amplitude, IMF 5 might contain useful information of inter-area oscillation. IMF 6 to IMF 9 are the frequency trends that features constant relative phase angles for all measurements. Fig. 12 shows the joint instantaneous frequency based on Hilbert spectral analysis. To identify the dominant oscillation mode, Fig. 13 shows the energy distribution of IMF1 to IMF 5, which have oscillatory phases among all IMFs. It can be seen that IMF 5 has the largest oscillation energy compared with other IMFs, indicating it is the dominant oscillation mode. Fig. 14 shows that MEMD can discover the dominant oscillation mode from both ambient and event measurements.

To further analyze these two oscillation components, Fig. 15 present the frequency and the mode compass graphs of the two oscillations. It shows that the two oscillations in ambient and event, both of which can be identified by MEMD, actually belong to the same inter-area oscillation mode.

For comparison, three typical methods: the modified Yule Walker method, the Prony's method, and the Fast

Fourier Transform (FFT) method are tested using the same measurements.

Table 6 shows their results, as well as a comparison on their advantages and disadvantages. As an example. The single-sided amplitude spectrum of the raw frequency measurement obtained by FFT is presented in Fig. 16, which displays a crest near 0.2 Hz indicating this oscillation mode. However, FFT does not preserve localized frequency, magnitude, as well as phase angle information over the studied time window.

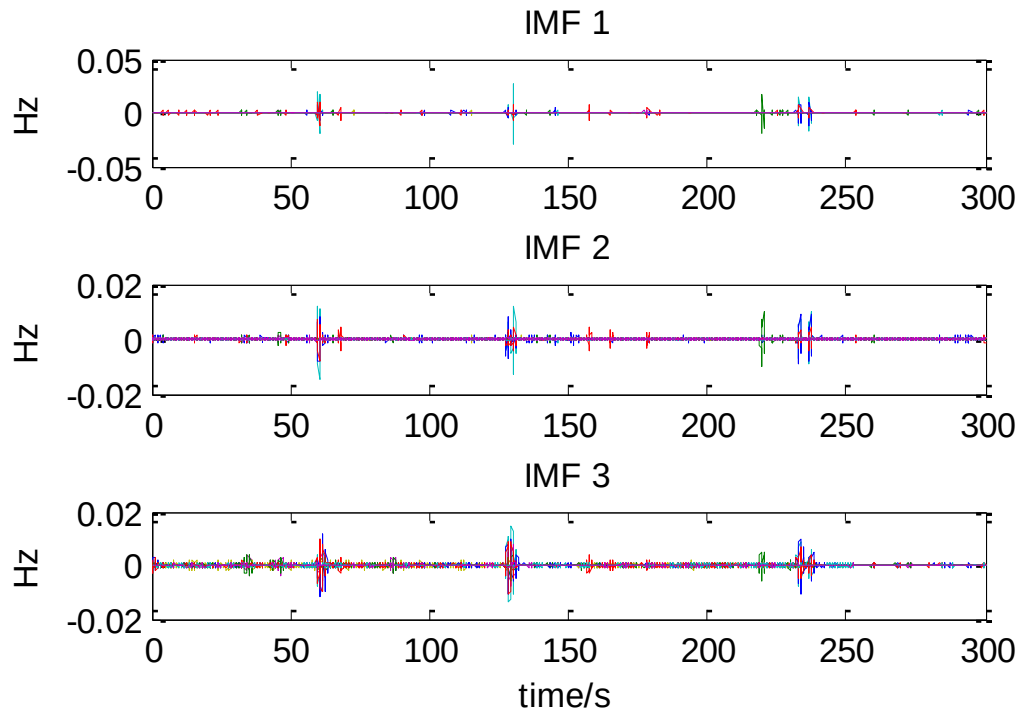

Fig. 9. IMF1-3 of the EI generation trip event

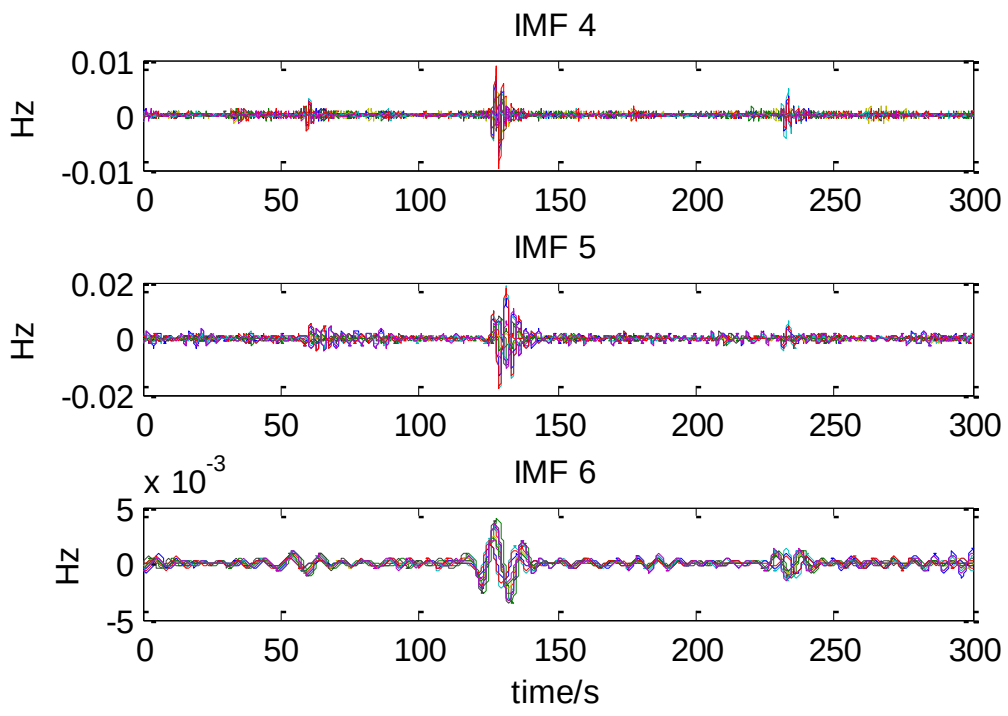

Fig. 10. IMF4-6 of the EI generation trip event

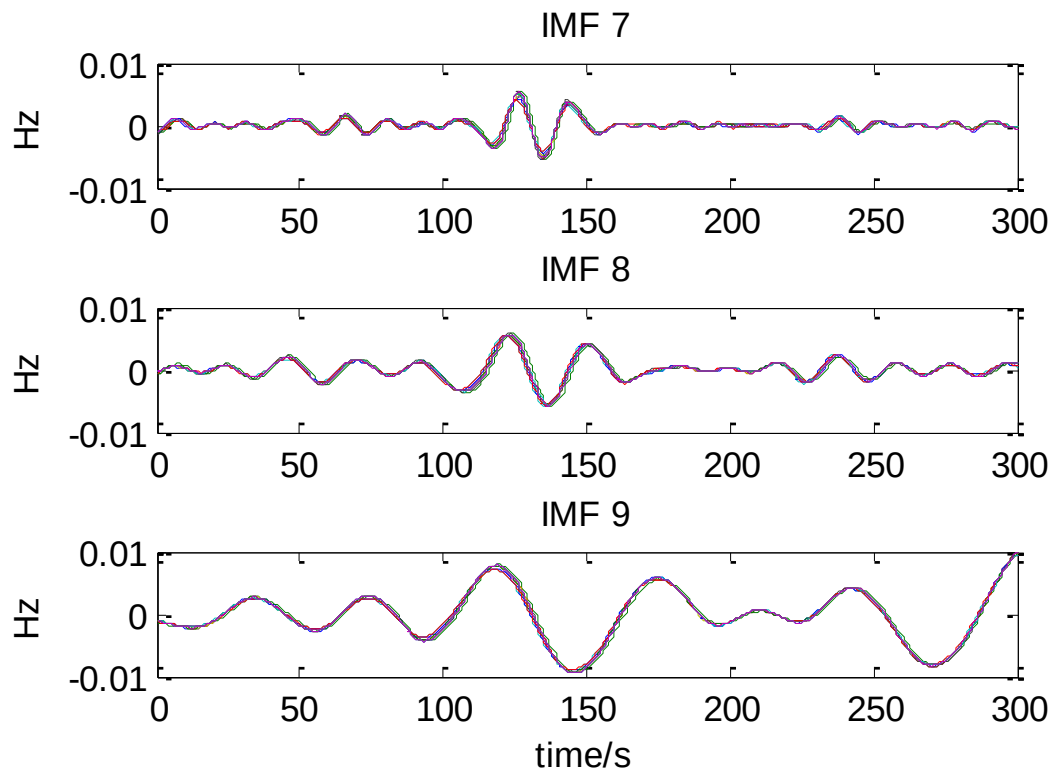

Fig. 11. IMF7-9 of the EI generation trip event

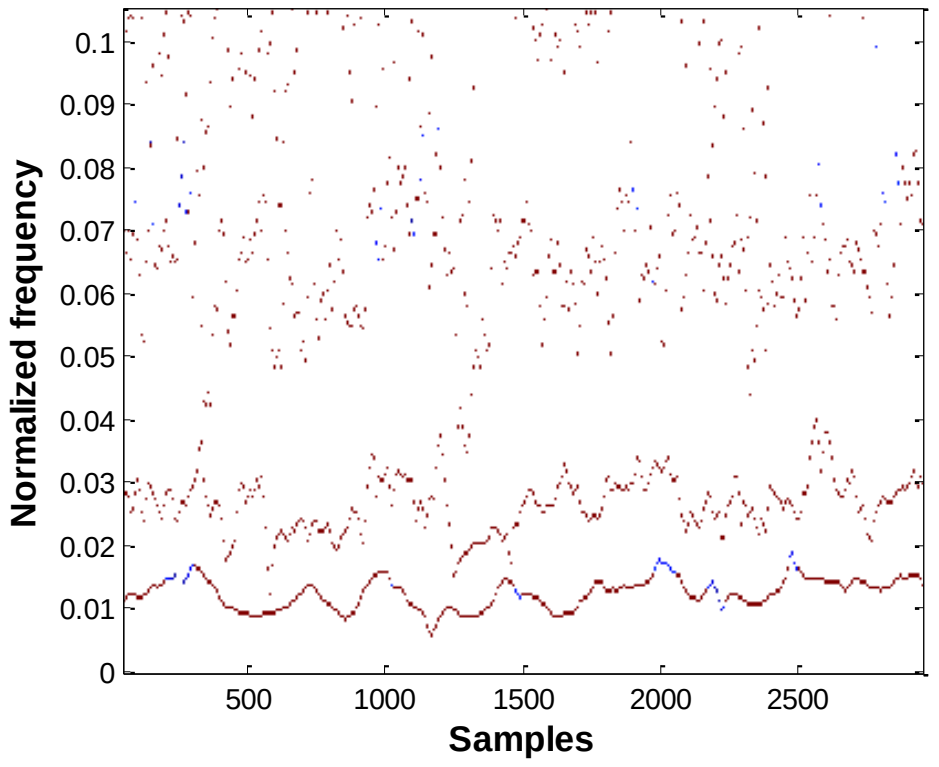

Fig. 12. The Hilbert spectrum (instantons frequency) of IMF 3 to IMF 5 in the EI generation trip event

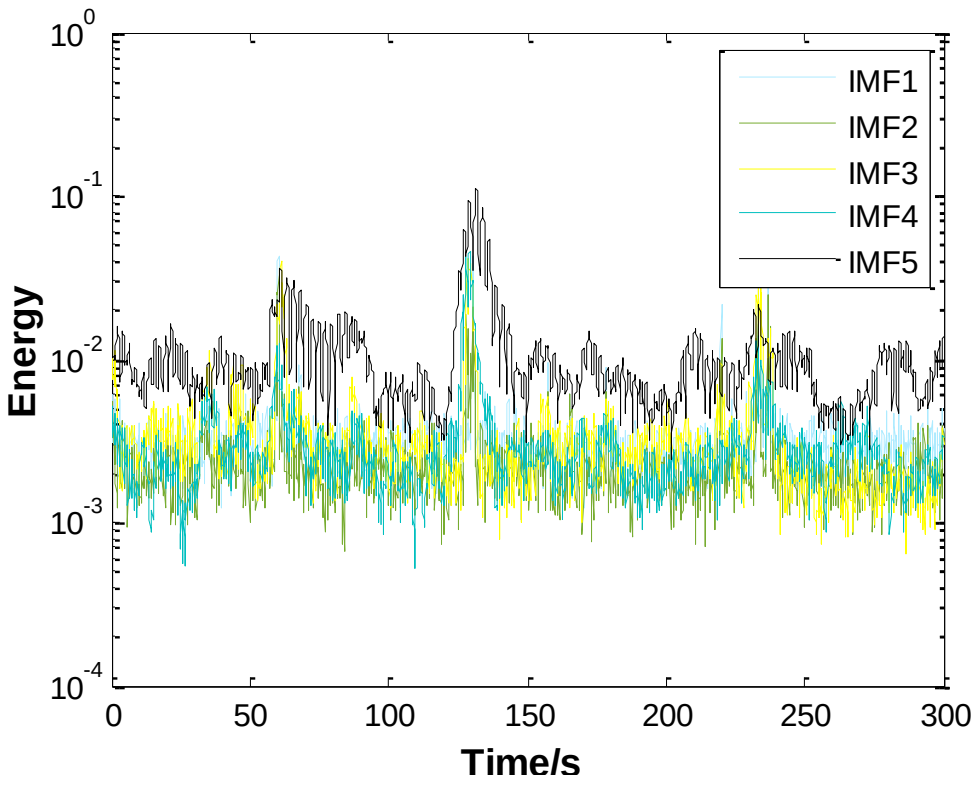

Fig. 13. The energy of IMF1-5 of the EI generation trip event

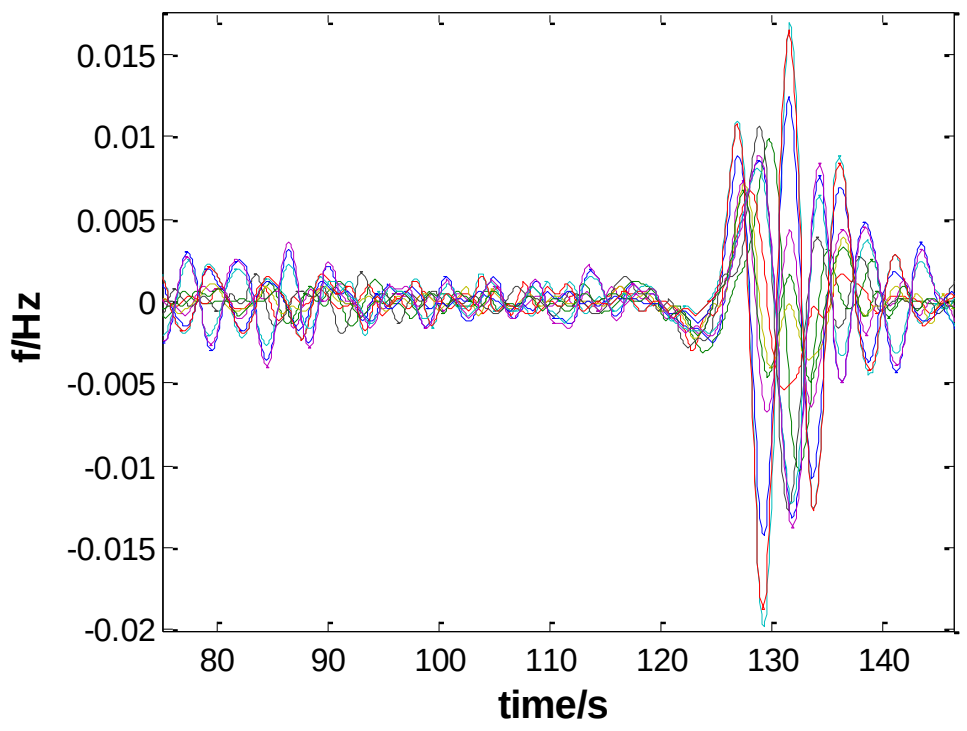

Fig. 14. IMF 5 of the EI generation trip event

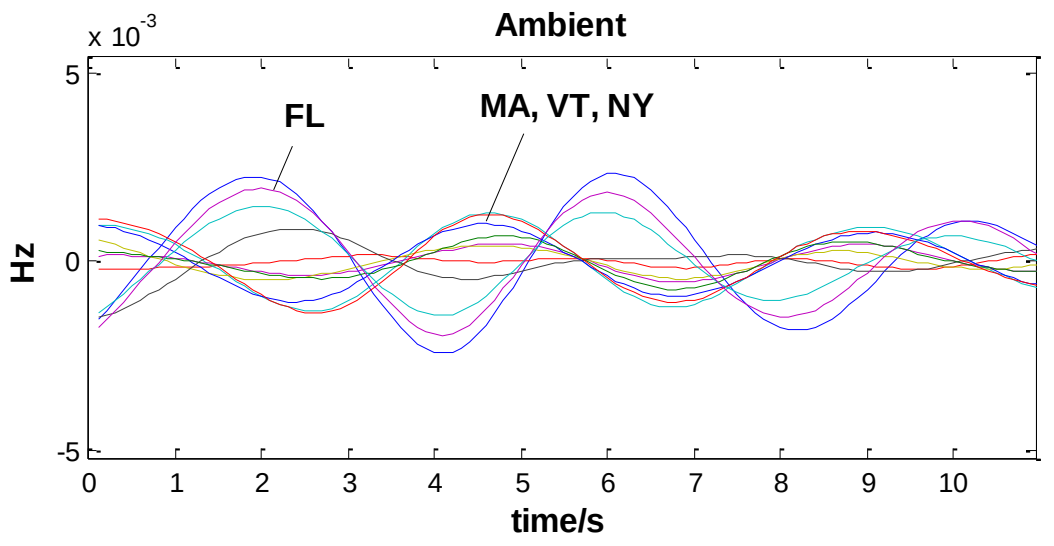

(a) Ambient measurements

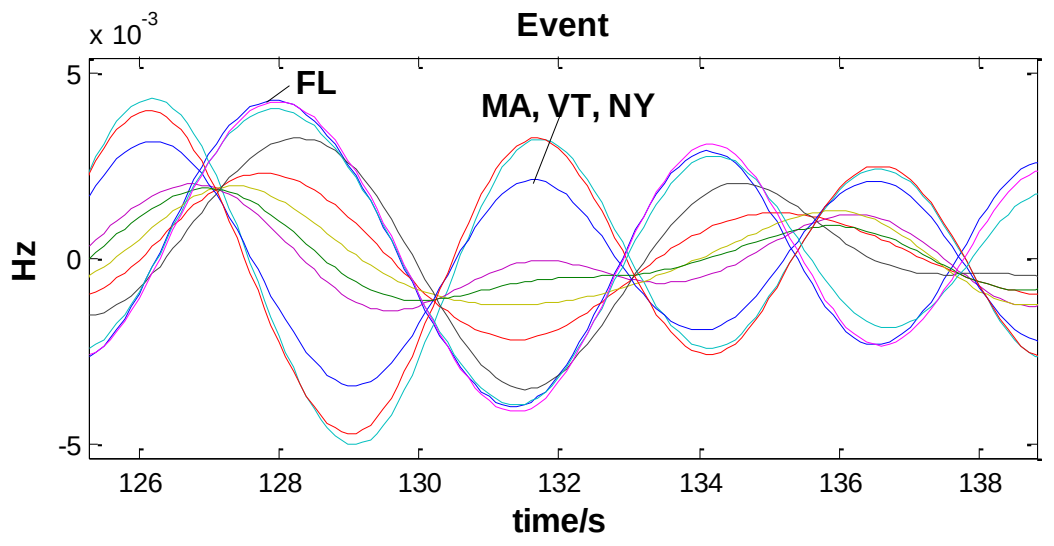

(b) Event measurements

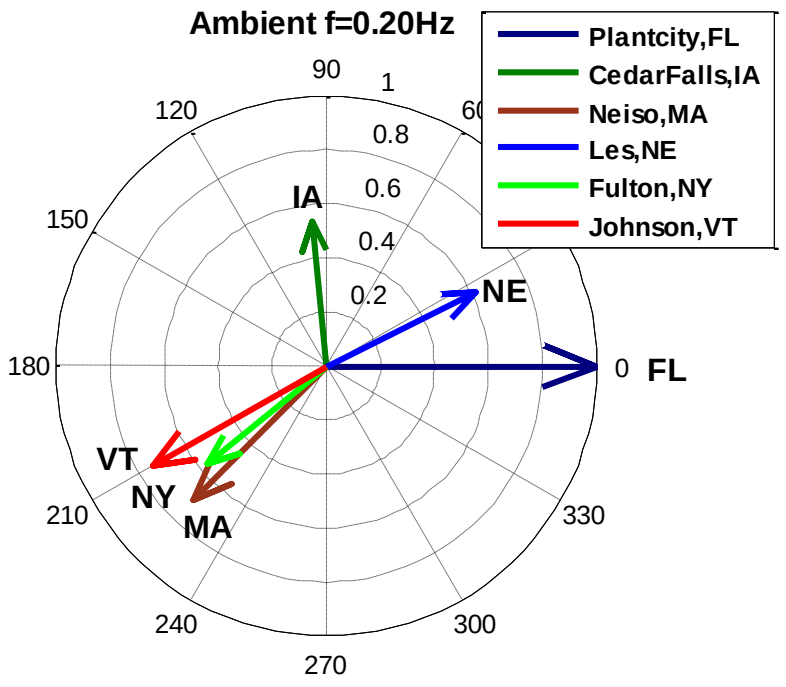

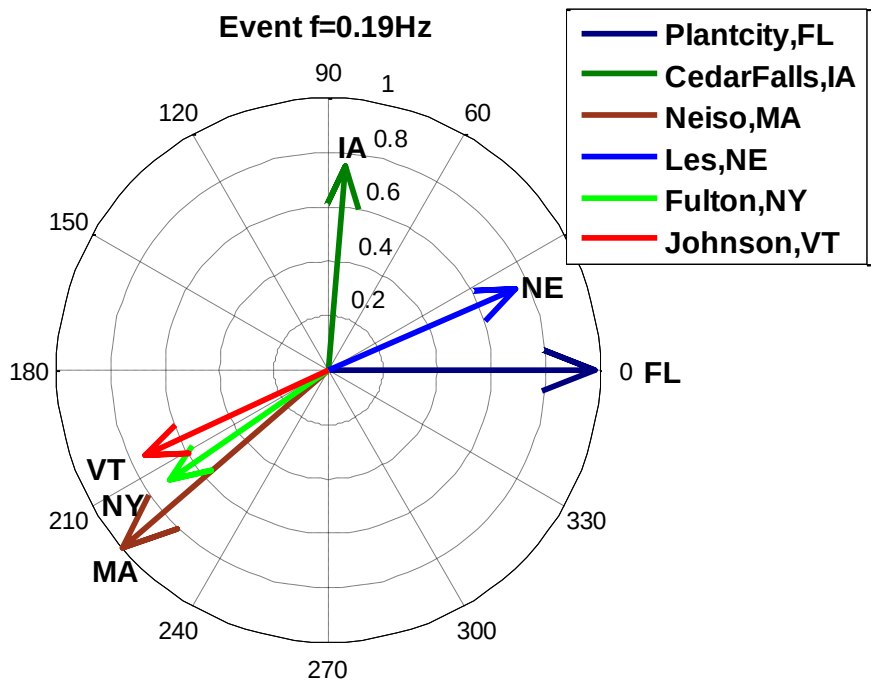

(c) Ambient oscillation mode compass plot

(d) Event oscillation mode compass plot

Fig. 15. Comparison on the identified oscillation mode based on EI ambient and event measurements (IMF 5)

Table 6. Method comparison based on EI frequency measurements

| Methods | Oscillation frequency | Advantage(s) | Disadvantage(s) |
|---|---|---|---|
| MEMD | Ambient: 0.20 Hz<br>Event: 0.19 HZ | a) Capable of analyzing non-stationary signals;<br>b) Provide localized frequency and amplitude;<br>c) Multiple signal analysis capability;<br>d) Preserve phase information; etc. | a) As a data-driven method, it needs more theoretical research.[37] |
| Fast Fourier Transform [67] | Ambient: 0.20 Hz<br>Event: 0.19 HZ | a) Computationally efficient;<br>b) Accurate frequency domain results;<br>c) Capable of analyzing non-stationary signals. | a) Lost information on localized frequency, amplitude, and phase when analyzing non-stationary signals |
| The modified Yule Walker method [14, 19, 68] | Ambient: 0.21 Hz<br>Event: 0.20 HZ | a) Capable of analyzing non-stationary signals;<br>b) As a parametric method, it has complete theoretical support. | a) Need pre-processing, such as de-trending;<br>b) Lost localized frequency and phase information. |
| The Prony's method [69] | N/A (due to non-stationary and noisy signals) | a) Computationally efficient;<br>b) As a parametric method, it has complete theoretical support. | a) Require the signal to be stationary. |

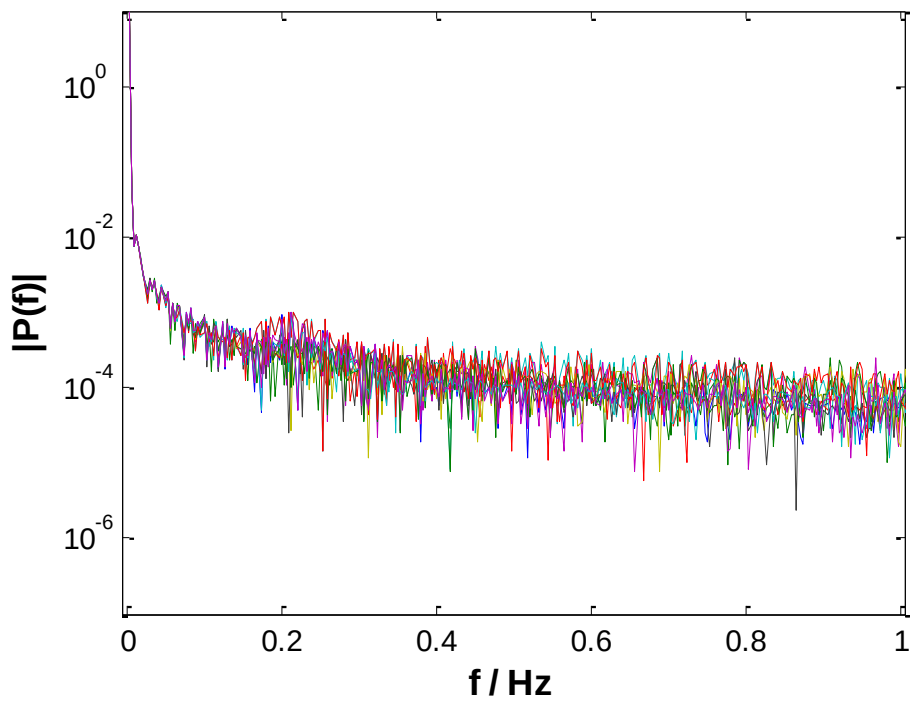

Fig. 16. FFT transform of raw frequency measurements of the EI generation trip event

## 5. Conclusions

This paper introduced BEMD, TEMD, and MEMD as multi-channel data-drive approaches for inter-area oscillation identification based on wide-area ambient synchrophasor measurements. The proposed method has the capability to identify inter-area oscillation modes using highly noisy ambient measurements. Moreover, it is robust to drifting, non-linear and non-stationary measurement signals. Test results based on real-world frequency measurements and comparison with existing methods show that the proposed methods have good potential for real-time monitoring and identification of inter-area oscillation modes.

When the measurement in one area is totally unavailable, the oscillation mode information of that area, such as the oscillation amplitude and phase angle, could not be re-constructed using MEMD as it is a measurement-based approach. Future work could be model-measurement hybrid oscillation identification method development and the optimal measurements selection for MEMD analysis in large power systems.